\documentclass[12pt]{article}
\usepackage{graphicx}
\usepackage{multirow}
\usepackage{subfigure}
\input epsf

\usepackage{subfigure}
\usepackage[table]{xcolor}
\usepackage{colortbl}
\definecolor{lightgray}{gray}{0.9}
\usepackage{graphicx}
\usepackage{lscape}
\usepackage{citesort}
\usepackage{amssymb}
\usepackage{appendix}
\usepackage{multirow}

\usepackage{graphicx}
\usepackage{lscape}
\usepackage{citesort}
\usepackage{amssymb}
\usepackage{appendix}
\usepackage{multirow}

\usepackage{mathrsfs}
\begin{document}
\def\qq{\langle \bar q q \rangle}
\def\uu{\langle \bar u u \rangle}
\def\dd{\langle \bar d d \rangle}
\def\sp{\langle \bar s s \rangle}
\def\GG{\langle g_s^2 G^2 \rangle}
\def\Tr{\mbox{Tr}}
\def\figt#1#2#3{
        \begin{figure}
        $\left. \right.$
        \vspace*{-2cm}
        \begin{center}
        \includegraphics[width=10cm]{#1}
        \end{center}
        \vspace*{-0.2cm}
        \caption{#3}
        \label{#2}
        \end{figure}
    }

\def\figb#1#2#3{
        \begin{figure}
        $\left. \right.$
        \vspace*{-1cm}
        \begin{center}
        \includegraphics[width=10cm]{#1}
        \end{center}
        \vspace*{-0.2cm}
        \caption{#3}
        \label{#2}
        \end{figure}
                }

\def\ds{\displaystyle}
\def\beq{\begin{equation}}
\def\eeq{\end{equation}}
\def\bea{\begin{eqnarray}}
\def\eea{\end{eqnarray}}
\def\beeq{\begin{eqnarray}}
\def\eeeq{\end{eqnarray}}
\def\ve{\vert}
\def\vel{\left|}
\def\ver{\right|}
\def\nnb{\nonumber}
\def\ga{\left(}
\def\dr{\right)}
\def\aga{\left\{}
\def\adr{\right\}}
\def\lla{\left<}
\def\rra{\right>}
\def\rar{\rightarrow}
\def\lrar{\leftrightarrow}
\def\nnb{\nonumber}
\def\la{\langle}
\def\ra{\rangle}
\def\ba{\begin{array}}
\def\ea{\end{array}}
\def\tr{\mbox{Tr}}
\def\ssp{{\Sigma^{*+}}}
\def\sso{{\Sigma^{*0}}}
\def\ssm{{\Sigma^{*-}}}
\def\xis0{{\Xi^{*0}}}
\def\xism{{\Xi^{*-}}}
\def\qs{\la \bar s s \ra}
\def\qu{\la \bar u u \ra}
\def\qd{\la \bar d d \ra}
\def\qq{\la \bar q q \ra}
\def\gGgG{\la g^2 G^2 \ra}
\def\q{\gamma_5 \not\!q}
\def\x{\gamma_5 \not\!x}
\def\g5{\gamma_5}
\def\sb{S_Q^{cf}}
\def\sd{S_d^{be}}
\def\su{S_u^{ad}}
\def\sbp{{S}_Q^{'cf}}
\def\sdp{{S}_d^{'be}}
\def\sup{{S}_u^{'ad}}
\def\ssp{{S}_s^{'??}}

\def\sig{\sigma_{\mu \nu} \gamma_5 p^\mu q^\nu}
\def\fo{f_0(\frac{s_0}{M^2})}
\def\ffi{f_1(\frac{s_0}{M^2})}
\def\fii{f_2(\frac{s_0}{M^2})}
\def\O{{\cal O}}
\def\sl{{\Sigma^0 \Lambda}}
\def\es{\!\!\! &=& \!\!\!}
\def\ap{\!\!\! &\approx& \!\!\!}
\def\md{\!\!\!\! &\mid& \!\!\!\!}
\def\ar{&+& \!\!\!}
\def\ek{&-& \!\!\!}
\def\kek{\!\!\!&-& \!\!\!}
\def\cp{&\times& \!\!\!}
\def\se{\!\!\! &\simeq& \!\!\!}
\def\eqv{&\equiv& \!\!\!}
\def\kpm{&\pm& \!\!\!}
\def\kmp{&\mp& \!\!\!}
\def\mcdot{\!\cdot\!}
\def\erar{&\rightarrow&}
\def\olra{\stackrel{\leftrightarrow}}
\def\ola{\stackrel{\leftarrow}}
\def\ora{\stackrel{\rightarrow}}

\def\simlt{\stackrel{<}{{}_\sim}}
\def\simgt{\stackrel{>}{{}_\sim}}


\title{
         {\Large
                 {\bf
                   Mass and magnetic dipole moment of   negative parity heavy baryons with spin--3/2
                 }
         }
      }

\author{
   K. Azizi\thanks {e-mail:kazizi@dogus.edu.tr, kazem.azizi@cern.ch},
  H. Sundu \thanks {e-mail:hayriye.sundu@kocaeli.edu.tr} \\
   \small$\ast$ Department of Physics, Do\u gu\c s University, Ac{\i}badem-Kad{\i}k\"oy, 34722
  Istanbul, Turkey \\
  \small $\dag$ Department of Physics, Kocaeli University, 41380
Izmit, Turkey}
 \date{}

\begin{titlepage}
\maketitle
\thispagestyle{empty}

\begin{abstract}
We calculate the mass and residue of the heavy  spin--3/2 negative parity baryons with single heavy bottom or charm quark
by the help of a two-point correlation function.
 We use the obtained results  to investigate the diagonal radiative transitions among the baryons under consideration. In particular, we compute corresponding transition form factors
via light cone QCD sum rules which are then used to obtain the magnetic dipole moments of the  heavy  spin--3/2 negative parity baryons. We remove the
pollutions coming from the positive parity spin--3/2 and positive/negative parity spin--1/2 baryons by constructing  sum rules for different Lorentz structures. We compare the results obtained with the existing theoretical
predictions.

\end{abstract}

~~~PACS number(s): 13.40.Gp, 13.40.Em, 14.20.Lq, 14.20.Mr, 11.55.Hx
\end{titlepage}

\section{Introduction}

The investigation of electromagnetic properties of hadrons including their electromagnetic form factors and multipole moments   is one of the key tools in understanding their internal structure
 and geometric shapes. Such properties of positive parity hadrons for both the light and heavy systems have been widely studied theoretically. From the experimental side, the subject of nucleon
electromagnetic form factors have been in the focus of much attention for many years although the experimental data via different experiments on some form factors are not in a 
good agreement (see for instance \cite{bra,Jefferson,Jones,Gayou1,Gayou2,Walker,Andivahis,Litt,Berger,Janssens,Arrington,Christy,Qattan}). We hope we will overcome these deficiencies and
experimentally study the electromagnetic  parameters of other light and heavy baryons by the developments in the facility of different experiments. However, we have even little
 theoretical  knowledge on the multipole  moments of the negative parity baryons. Hence, both theoretical and experimental studies on the electromagnetic form factors of negative parity states are welcome as they
can help us gain valuable knowledge on the nature of the strong interactions inside these objects.   Jefferson Laboratory and Mainz
Microton facility  are planing to measure the electromagnetic form factors and multipole moments of the negative parity baryons  \cite{Jefferson,Mainz3}.
 It is desired that the new electron beam facilities would  allow to compile a large
number of more precise  data to study the electro-excitations of the nucleon resonances.

The theoretical studies on the spectroscopic and electromagnetic decays of negative parity baryons have mainly been devoted to the radial excitations of the nucleons and other light baryons (see for instance \cite{Chung,Oka,Kondo,Aliev,savci3,hayriye} and references therein).  In the heavy sector, some spectroscopic properties of negative parity  spin--1/2 and spin--3/2 heavy baryons  have been studied
in \cite{Zhi-Gang Wang1,Zhi-Gang Wang2,Roberts,Ebert1,Ebert2}. The magnetic moments of negative parity heavy baryons with $J^P = {1\over 2}^-$ are calculated in \cite{savci4}.
The transition magnetic moments between negative parity heavy spin--1/2 baryons are also investigated in \cite{savci5,savci6}.
In the present work, first we study the masses and  residues  of the heavy spin--3/2 negative parity baryons using a two-point correlation function in the context of QCD or SVZ sum rules \cite{Shifman}.
The obtained results are then used to compute the magnetic dipole moments of the heavy spin--3/2 negative parity baryons in the frameworks of   light cone QCD sum rules (LCSR) by the help of photon distribution amplitudes
(DAs). A similar calculations on the spectroscopic and electromagnetic properties of the  heavy spin--3/2 positive parity baryons can be found in \cite{altug}.
The interpolating current of the baryons under consideration in the present study also couple to the  heavy spin--3/2 positive parity baryons as well as heavy spin--1/2 baryons with
both parities. To remove the unwanted contributions coming from these channels, different Lorentz structures entering the calculations as well as an appropriate ordering of the Dirac matrices are used.

The paper is organized as follows. In next section, we construct a mass sum rule to evaluate the masses and pole residues of the negative parity heavy spin--3/2 baryons and numerically analyze
the obtained sum rules. The results are compared with the existing theoretical predictions. In section 3, we construct the  LCSR for the  electromagnetic form factors defining the diagonal radiative transitions
of the baryons under consideration and compute the corresponding magnetic dipole moments and their numerical values. The last section
 is dedicated to the concluding remarks.

\section{ Mass and residue of the negative parity spin--3/2 heavy baryons}

In order to calculate the mass and residue of the negative parity
spin-3/2 baryons with single heavy bottom or charm quark, we
start with the following two point correlation function:
\begin{eqnarray}\label{CorrelationFunctionTwoPoint}
\mathrm{T}_{\mu\nu}=i \int d^4x~e^{ik\cdot x}~ {\langle}0| {\cal
T}\left \{ \eta_{\mu}^{{\cal B}}(x)~\bar{\eta}_{\nu}^{{\cal
B}}(0)\right\}|0{\rangle},
\end{eqnarray}
where $\eta_{\mu}^{{\cal B}}$ is the interpolating current of ${\cal B}$ baryon coupling to both the positive and negative parity baryons.
To construct the mass sum rules for the baryons under consideration we calculate this correlation function via two different ways:
in terms of hadronic parameters and in terms of QCD degrees of freedom by the help of operator product expansion (OPE). The hadronic side of the correlation
function  is
obtained by inserting complete sets of intermediate states with both parities. After performing the four-integral we get
\begin{eqnarray}\label{CorrelationFunctionTwoPoint}
\mathrm{T}_{\mu\nu}^{HAD}&=&\frac{{\langle}0|\eta_{\mu}^{{\cal B}}|{\cal{B}}^+(k,s){\rangle}
{\langle}{\cal{B}}^+(k,s)|\bar{\eta}_{\nu}^{{\cal B}}|0{\rangle}}{m^2_{{\cal{B}}^+}-k^2}
+\frac{{\langle}0|\eta_{\mu}^{{\cal B}}|{\cal{B}}^-(k,s){\rangle}
{\langle}{\cal{B}}^-(k,s)|\bar{\eta}_{\nu}^{{\cal B}}|0{\rangle}}{m^2_{{\cal{B}}^-}-k^2}
+...,
\end{eqnarray}
where  ${\cal{B}}^+$ and ${\cal{B}}^-$ denote the positive and negative parity spin-$3/2$ baryons,
respectively and $...$ shows the contributions of the higher states and continuum. To proceed, we need to define the following  matrix elements:
\begin{eqnarray}\label{MatrixElements}
{\langle}0|\eta_{\mu}^{{\cal B}}|{\cal B}^+(k,s){\rangle}&=&\lambda_{{\cal
B}^+}u_{\mu}(k,s),
\nonumber \\
{\langle}0|\eta_{\mu}^{{\cal B}}|{\cal B}^-(k,s){\rangle}&=&\lambda_{{\cal
B}^-}\gamma_5 u_{\mu}(k,s),
\end{eqnarray}
where $u_{\mu}(k,s)$ is the Rarita-Schwinger spinor for the
spin-$3/2$ particles and $\lambda_{{\cal
B}^{\pm}}$ are the residues of the ${\cal{B}}^{\pm}$ baryons. Here we shall comment that the $\eta_{\mu}^{{\cal B}}$ current not only interacts with the spin--3/2 states, but also with the
spin--1/2 states. Hence, we should remove the unwanted pollution coming from the  spin--1/2 states.
The general form of the matrix element of $\eta_{\mu}^{{\cal B}}$ between the spin-1/2 and vacuum
states can be written as
\begin{eqnarray}\label{EliminationSpin1half2}
{\langle}0|\eta_{\mu}^{{\cal B}}|\frac{1}{2}(k){\rangle}=\left(A k_{\mu}+B
\gamma_{\mu}\right)u(k),
\end{eqnarray}
where $A$ and $B$ are some constants.
By multiplication of both sides of this equation with $\gamma^\mu$, and by using the
condition $\gamma^\mu \eta_{\mu}^{{\cal B}}=0$ to get rid of transverse spin--1/2 components (for details see \cite{Savvidy}), we get
\begin{eqnarray}\label{EliminationSpin1half2-1}
{\langle}0|\eta_{\mu}^{{\cal B}}|\frac{1}{2}^+(k){\rangle}=B\left(-\frac{4}{m_{\frac{1}{2}^+}}
k_{\mu}+ \gamma_{\mu}\right)u(k),
\end{eqnarray}
for the positive parity and
\begin{eqnarray}\label{EliminationSpin1half2-2}
{\langle}0|\eta_{\mu}^{{\cal B}}|\frac{1}{2}^-(k){\rangle}=B\gamma_{5}\left(-\frac{4}{m_{\frac{1}{2}^-}}
k_{\mu}+ \gamma_{\mu}\right)u(k),
\end{eqnarray}
for the negative parity states.
From these equations we see  that the unwanted contributions coming from the spin-1/2 states
 are proportional to either $k_\mu$ or $\gamma_\mu$.
 To remove these contributions, first we order the  Dirac matrices as $\gamma_\mu \rlap/{k}
 \gamma_\nu$ and then set the terms with $\gamma_\mu$ in the beginning and $\gamma_\nu$ at the end and those terms proportional to $k_\mu$ and $k_\nu$ to zero (for further details about the eliminating the 
contributions of the spin-1/2 particles see \cite{Aliev1}).

Now, we insert
Eq.~(\ref{MatrixElements}) into
Eq.~(\ref{CorrelationFunctionTwoPoint}) and use the relation
$\mathrm{T}_{\mu\nu}^-=-\gamma_5\mathrm{T}_{\mu\nu}^+\gamma_5$
(see also \cite{Oka})  and the summation over the  spin-$3/2$ baryons
via
\begin{eqnarray}\label{SpinSummation}
\sum_s u_\mu (k,s) \bar{u}_\nu (k,s)
=-(\not\!k+m_{\cal{B}})\Big[g_{\mu\nu}-\frac{1}{3}\gamma_{\mu}\gamma_{\nu}
-\frac{2k_{\mu}k_{\nu}}{3m_{{\cal {
B}}}^2}+\frac{k_{\mu}\gamma_{\nu}-k_{\nu}\gamma_{\mu}}{3m_{{\cal
{B}}}}\Big].
\end{eqnarray}
As a result, for the hadronic side of the correlation function in
the Borel scheme, we get
\begin{eqnarray}\label{CorrelationFunctionTwoPoint1HAD}
\widehat{\textbf{B}}\mathrm{T}_{\mu\nu}^{HAD}&=&
-\Big[\lambda_{{\cal{B^+}}}^2e^{-\frac{m_{{\cal{B^+}}}^2}{M_B^2}}
+\lambda_{{\cal{B^-}}}^2e^{-\frac{m_{{\cal{B^-}}}^2}{M_B^2}}\Big]
\not\!k
g_{\mu\nu}-\Big[\lambda_{{\cal{B^+}}}^2m_{{\cal{B^+}}}e^{-\frac{m_{{\cal{B^+}}}^2}{M_B^2}}
-\lambda_{{\cal{B^-}}}^2m_{{\cal{B^-}}}e^{-\frac{m_{{\cal{B^-}}}^2}{M_B^2}}\Big]g_{\mu\nu}
\nonumber \\
&+& \mbox{other structures},
\end{eqnarray}
where $M_B^2$ is the Borel mass parameter coming from the Borel transformation which is performed to suppress the contributions of the higher states and continuum.

The OPE side of the aforesaid correlation function is calculated  in terms
of the QCD degrees of freedom in deep Euclidean region. To this aim, we need the explicit form of the
interpolating current  $\eta_{\mu}^{{\cal B}}$  which is given as (for some details about the baryon currents see \cite{Chung,Ioffe,Dosch,Lee})
\begin{eqnarray}\label{currents}
\eta_{\mu}^{{\cal B}}=A\epsilon_{ijk}\Big\{(q_{1}^{iT}C\gamma_{\mu}q_{2}^{j})Q^{k}+(q_{2}^{iT}C\gamma_{\mu}Q^{j})q_{1}^{k}+
(Q^{iT}C\gamma_{\mu}q_{1}^{j})q_{2}^{k}\Big\},
\end{eqnarray}
where $C$ is the charge conjugation operator;  $i$, $j$ and $k$
are color indices and $Q$ denotes the heavy $b$ or $c$ quark. The normalization constant $A$ and the light quark fields $q_1$ and
$q_2$ for each heavy baryon is given in table 1.
\begin{table}[h]
\centering
\begin{tabular}{|c||c|c|c|}\hline
 &A & $q_{1}$& $q_{2}$\\\cline{1-4}
\hline\hline
 $\Sigma_{b(c)}^{*+(++)}$&$1/\sqrt{3}$ &u&u\\\cline{1-4}
 $\Sigma_{b(c)}^{*0(+)}$&$\sqrt{2/3}$&u&d\\\cline{1-4}
 $\Sigma_{b(c)}^{*-(0)}$&$1/\sqrt{3}$&d&d\\\cline{1-4}
 $\Xi_{b(c)}^{*0(+)}$&$\sqrt{2/3}$&s&u\\\cline{1-4}
 $\Xi_{b(c)}^{*-(0)}$&$\sqrt{2/3}$&s&d\\\cline{1-4}
$\Omega_{b(c)}^{*-(0)}$ &$1/\sqrt{3}$&s &s\\\cline{1-4}
 \end{tabular}
 \vspace{0.8cm}
\caption{The normalization constant $A$ and the light quark fields $q_{1}$ and $q_{2}$  for
the corresponding baryons. }\label{tab:2}
\end{table}

After inserting the  explicit form of the interpolating current
into the correlation function in
Eq.~~(\ref{CorrelationFunctionTwoPoint}) and performing
contractions via the Wick's theorem, we get the OPE side in terms of the heavy and light quarks propagators.
For the light quark
propagator in coordinate space we use \cite {Balitsky}
\bea
\label{eh32v18}
S_q(x) \es {i \rlap/x\over 2\pi^2 x^4} - {m_q\over 4 \pi^2 x^2} -
{\lla \bar q q \rra\over 12} \left(1 - i {m_q\over 4} \rlap/x \right) -
{x^2\over 192} m_0^2 \lla \bar q q \rra  \left( 1 -
i {m_q\over 6}\rlap/x \right) \nnb \\
&&  - i g_s \int_0^1 du \left[{\rlap/x\over 16 \pi^2 x^2} G_{\mu \nu} (ux)
\sigma_{\mu \nu} - {i\over 4 \pi^2 x^2} u x^\mu G_{\mu \nu} (ux) \gamma^\nu
\right. \nnb \\
&& \left.
 - i {m_q\over 32 \pi^2} G_{\mu \nu} \sigma^{\mu
 \nu} \left( \ln \left( {-x^2 \Lambda^2\over 4} \right) +
 2 \gamma_E \right) \right]~,
\eea
where $\gamma_E \simeq 0.577$ is the Euler constant and $\Lambda$ is a
scale parameter.
The heavy quark propagator in an external field is also taken as
\bea
\label{eh32v19}
S_Q(x) = S_Q^{free}(x) -
ig_s \int {d^4k \over (2\pi)^4} e^{-ikx} \int_0^1
du \Bigg[ {\rlap/k+m_Q \over 2 (m_Q^2-k^2)^2} G^{\mu\nu} (ux)
\sigma_{\mu\nu} +
{u \over m_Q^2-k^2} x_\mu G^{\mu\nu} \gamma_\nu \Bigg]~,\nonumber \\
\eea
where $S_Q^{free}(x)$ is the free heavy quark operator in x--representation and
is given by
\bea
\label{nolabel}
S_Q^{free} (x) =
{m_Q^2 \over 4 \pi^2} {K_1(m_Q\sqrt{-x^2}) \over \sqrt{-x^2}} -
i {m_Q^2 \rlap/{x} \over 4 \pi^2 x^2} K_2(m_Q\sqrt{-x^2})~,
\eea
where $K_1$ and $K_2$ are the modified Bessel function of the second kind.
By using these propagators in the coordinate space and performing the Fourier and Borel transformations as well
as applying the continuum subtraction, after a very lengthy calculations we get
\begin{eqnarray}\label{CorrelationFunctionTwoPointOPE}
\widehat{\textbf{B}}\mathrm{T}_{\mu\nu}^{OPE}&=&\mathrm{T}_1^{OPE}
\not\!k g_{\mu\nu}+\mathrm{T}_2^{OPE} g_{\mu\nu}+\mbox{other
structures},
\end{eqnarray}
where the functions $\mathrm{T}_{1,2}^{OPE}$, for instance  for $\Sigma_{b}^{*0}$ particle, are given as
\begin{eqnarray}\label{T1}
\mathrm{T}_1^{OPE}&=& \int_{m_b^2}^{s_0} ds
e^{-\frac{s}{M_{{B}}^2}}\Bigg\{\frac{15m_b^8s-4m_b^{10}+40m_b^4s^3-60m_b^2s^4+9s^5
+60m_b^4s^3\ln[\frac{s}{m_b^2}]}{15\times2^8\pi^4s^3}
\nonumber \\
&-&\Big({\langle}u\bar{u}{\rangle}+{\langle}d\bar{d}{\rangle}\Big)\frac{m_b(m_b^2-s)^2}{24\pi^2s^2}
-{\langle}g_s^2GG{\rangle}\frac{(m_b^6-4m_b^2s^2+3s^3)}{2^7\times
3^2\pi^4s^3}-{\langle}g_s^2GG{\rangle}^2\frac{m_b^2}{2^8\times3^4\pi^4s^3}
\nonumber \\
&+&m_0^2\Big({\langle}u\bar{u}{\rangle}+{\langle}d\bar{d}{\rangle}\Big)
\frac{m_b^3}{48\pi^2s^2}-{\langle}g_s^2GG{\rangle}\Big({\langle}u\bar{u}{\rangle}+{\langle}d\bar{d}{\rangle}\Big)
\frac{m_b}{2^5\times3^2\pi^2s^2}
 \Bigg\}
\nonumber \\
&+&e^{-\frac{m_b^2}{M_{{{B}}}^2}}\Bigg\{\frac{1}{9}{\langle}u\bar{u}{\rangle}{\langle}d\bar{d}{\rangle}+{\langle}g_s^2GG{\rangle}
\Big({\langle}u\bar{u}{\rangle}+{\langle}d\bar{d}{\rangle}\Big)\frac{1}{2^5\times3^3\pi^2m_b}
-m_0^2{\langle}u\bar{u}{\rangle}{\langle}d\bar{d}{\rangle}\frac{m_b^2+M_{{B}}^2}{18M_{{B}}^4}
\nonumber \\
&+&{\langle}g_s^2GG{\rangle}^2\frac{m_b^2+2M_{{B}}^2}{2^{10}\times3^4\pi^4m_b^2M_{{B}}^2}
-m_0^2
{\langle}d\bar{d}{\rangle}{\langle}g_s^2GG{\rangle}\frac{m_b^2-2M_{{B}}^2}{2^7\times3^3\pi^2m_bM_{{B}}^4}
\nonumber \\
&+&m_0^2{\langle}g_s^2GG{\rangle}{\langle}d\bar{d}{\rangle}{\langle}u\bar{u}{\rangle}
\frac{m_b^2(m_b^2-M_{{B}}^2)}{2^4\times3^4M_{{B}}^{10}}-{\langle}d\bar{d}{\rangle}{\langle}u\bar{u}{\rangle}
{\langle}g_s^2GG{\rangle}\frac{m_b^2}{2^3\times3^4M_{{B}}^{6}}
\nonumber \\
&-&m_0^2{\langle}g_s^2GG{\rangle}{\langle}u\bar{u}{\rangle}\frac{m_b^2-2M_{{B}}^2}
{2^7\times3^3\pi^2m_bM_{{B}}^4}\Bigg\},
\end{eqnarray}
and
\begin{eqnarray}\label{T2}
\mathrm{T}_2^{OPE}&=& \int_{m_b^2}^{s_0} ds
e^{-\frac{s}{M_{{{B}}}^2}}\Bigg\{\frac{(s-m_b^2)\Big(41m_b^4-sm_b^2+2s^2\Big)-6m_b^4
\Big(4s-3m_b^2\Big)\ln[\frac{s}{m_b^2}]}{2^7\times3^2
\pi^4 m_b}
\nonumber \\
&-&\frac{(4s+m_b^2)(m_b^2-s)^4}{2^8\times3^2\pi^4m_bs^2}-{\langle}u\bar{u}{\rangle}
\frac{(m_b^2-s)^2(m_b^4+2m_b^2s-6s^2)}{72\pi^2m_b^2s^2}-
{\langle}d\bar{d}{\rangle}
\frac{(m_b^2-s)^2(m_b^2+2s)}{72\pi^2s^2}
\nonumber \\
&+& {\langle}g_s^2GG{\rangle}\frac{24m_b^4s-11m_b^6
-21m_b^2s^2+8s^3+18m_b^2s^2\ln[\frac{m_b^2}{s}]}{2^9\times3^3\pi^4m_bs^2}
\nonumber \\
&+&\Big(m_0^2{\langle}u\bar{u}{\rangle}+m_0^2{\langle}d\bar{d}{\rangle}\Big)
\frac{(s^2+m_b^4)}{96\pi^2s^2}
-{\langle}g_s^2GG{\rangle}^2\frac{m_b}{2^{10}\times3^3\pi^4s}\Bigg\}
\nonumber \\
&+&e^{-\frac{m_b^2}{M_{{{B}}}^2}}\Bigg\{\frac{1}{3}m_b{\langle}u\bar{u}
{\rangle}{\langle}d\bar{d}{\rangle}-m_0^2{\langle}u\bar{u}
{\rangle}{\langle}d\bar{d}{\rangle}\frac{m_b^3}{6M_{{B}}^4}
+\frac{{\langle}g_s^2GG{\rangle}\Big({\langle}u\bar{u}
{\rangle}+{\langle}d\bar{d}{\rangle}\Big)}{2^4\times3^3 \pi^2}
\nonumber \\
&+&{\langle}g_s^2GG{\rangle}^2\frac{m_b^2-M_{{B}}^2}{2^{10}\times3^4\pi^4m_bM_{{B}}^2}
+m_0^2{\langle}g_s^2GG{\rangle}{\langle}u\bar{u}{\rangle}{\langle}d\bar{d}{\rangle}
\frac{m_b(m_b^4-6m_b^2M_{{B}}^2+6M_{{B}}^4)}{2^4\times3^3M_{{B}}^{10}}
\nonumber \\
&-&{\langle}g_s^2GG{\rangle}{\langle}u\bar{u}{\rangle}{\langle}d\bar{d}{\rangle}
\frac{m_b(m_b^2-3M_{{B}}^2)}{2^3\times3^3M_{{B}}^6}-m_0^2{\langle}g_s^2GG{\rangle}
\Big({\langle}u\bar{u}{\rangle}+{\langle}d\bar{d}{\rangle}\Big)\frac{m_b^2}{2^6\times3^2\pi^2M_{{B}}^4}
\Bigg\},
\end{eqnarray}
where $s_0$ is the continuum threshold, and for simplicity, we ignored to present the terms containing the light quark masses and those proportional to $m_0^4$. Note that we only ignored to present such terms
in the above formulas and we will take into account their contributions in the numerical calculations. 

Having calculated both the hadronic and OPE sides of the
correlation function, we match the coefficients of the structures $\not\!k
g_{\mu\nu}$ and $ g_{\mu\nu}$ from these two sides and obtain the mass and residue of the negative
parity heavy spin-$3/2$ baryons as
\begin{eqnarray}\label{MassResidue}
m_{{\cal{B^-}}}^2&=&\frac{\frac{d}{d\frac{1}{M_B^2}}(\mathrm{T}_2^{OPE}-m_{{\cal{B^+}}}
\mathrm{T}_1^{OPE})}{\mathrm{T}_2^{OPE}-m_{{\cal{B^+}}}\mathrm{T}_1^{OPE}},
\nonumber \\
\lambda_{{\cal{B^-}}}^2&=&\frac{\mathrm{T}_2^{OPE}-m_{{\cal{B^+}}}\mathrm{T}_1^{OPE}}
{m_{{\cal{B^+}}}+m_{{\cal{B^-}}}}e^{\frac{m_{{\cal{B^-}}}^2}{M_B^2}},
\end{eqnarray}
where $\frac{d}{d\frac{1}{M_B^2}}$ denotes derivative with respect to $\frac{1}{M_B^2}$.

To obtain the numerical values of the masses and residues of the negative parity heavy spin--3/2 baryons we take $\langle\bar{u}u\rangle(1~GeV)=\langle\bar{d}d\rangle(1~GeV)=-(0.24\pm0.01)^3 \mbox{GeV$^3$}$,
$\langle \bar{s}s\rangle(1~GeV)=0.8\langle \bar{u}u\rangle(1~GeV)~\mbox{GeV$^3$}$, $ \langle\frac{\alpha_sG^2}{\pi}\rangle =
(0.012\pm0.004)~\mbox{GeV$^4$}$ and $ m_0^2(1~GeV)=(0.8\pm0.2)$ $~\mbox{GeV$^2$}$ \cite{belyaev}. 
Beside these input parameters, we shall also find working regions of the auxiliary parameters $s_0$ and $M_B^2$ such that the physical quantities show weak
dependence on these parameters according to the standard criteria of the method used. The continuum threshold is not entirely arbitrary but it depends on the energy of the first
excited state with the same quantum numbers. Hence, according to the standard prescriptions, the value of this threshold is mainly chosen such that the mass of the first excited state remains above $\sqrt {s_0} $.   This may be considered as a weak point of the QCD sum rule approach especially in the case of the negative parity baryons. Since for many   states we know the exact values of the ground state masses experimentally but have not enough information on the energy of the corresponding first excited states. Considering this point, in our calculations, we impose the pole dominance condition and demand that the pole contribution consists the highest possible part of the whole result in each channel. This leads us to  take this parameter in the interval  $(m_{{\cal{B}}^-}+0.1)^2\leq s_0 \leq (m_{{\cal{B}}^-}+0.7)^2$.
\begin{figure}[h!]
\begin{center}
\subfigure[]{\includegraphics[width=8cm]{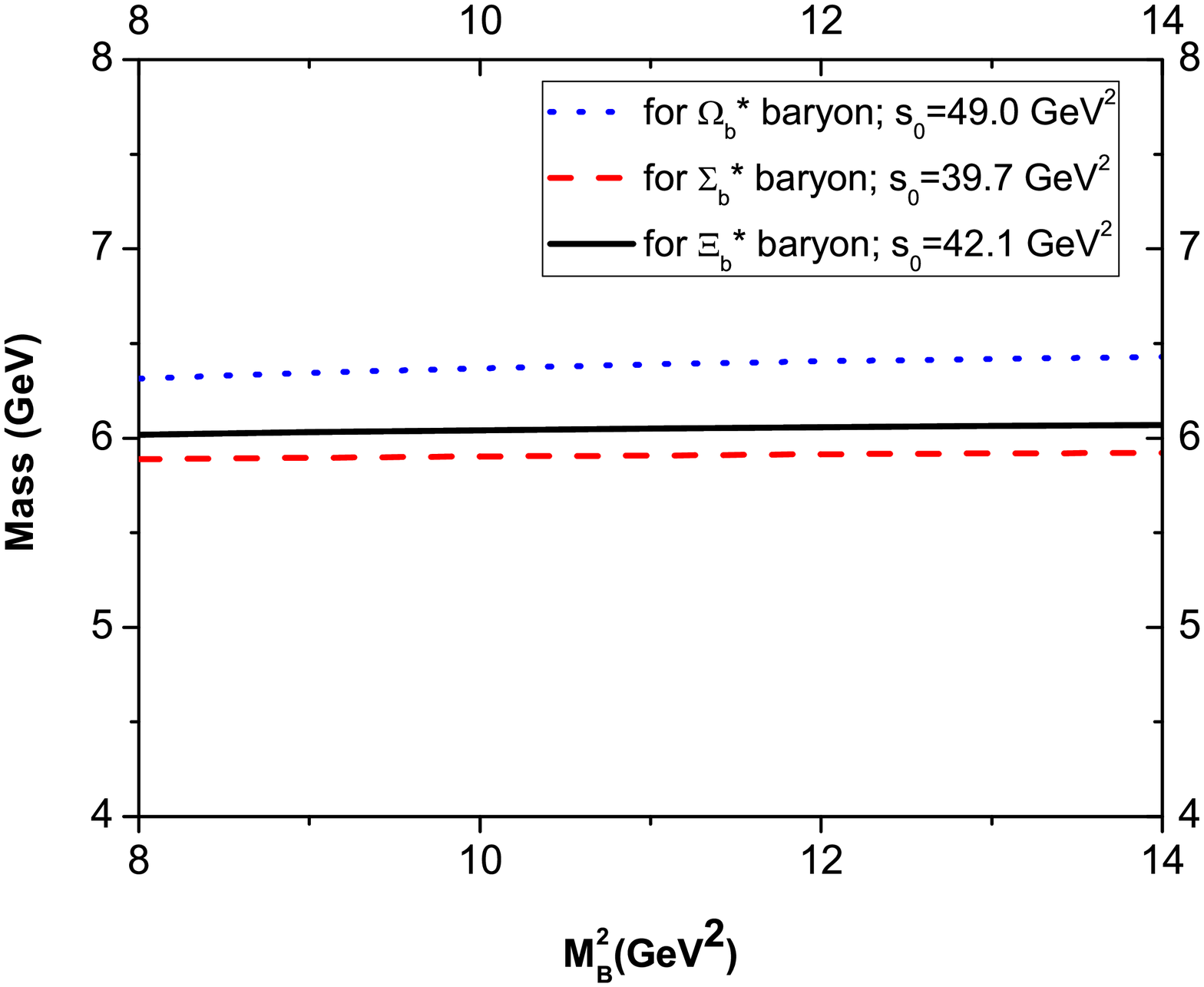}}
\subfigure[]{\includegraphics[width=8cm]{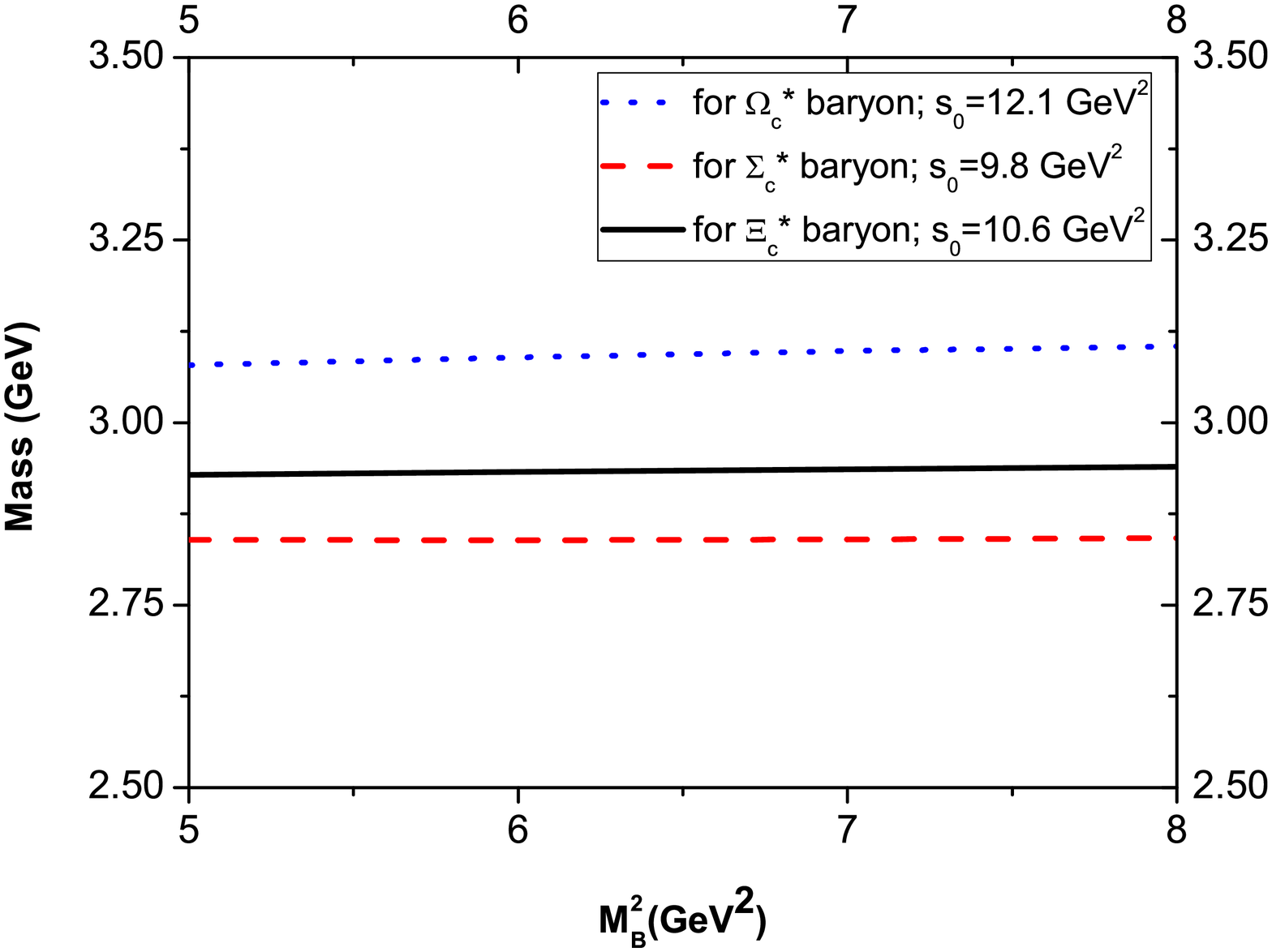}}
\subfigure[]{\includegraphics[width=8cm]{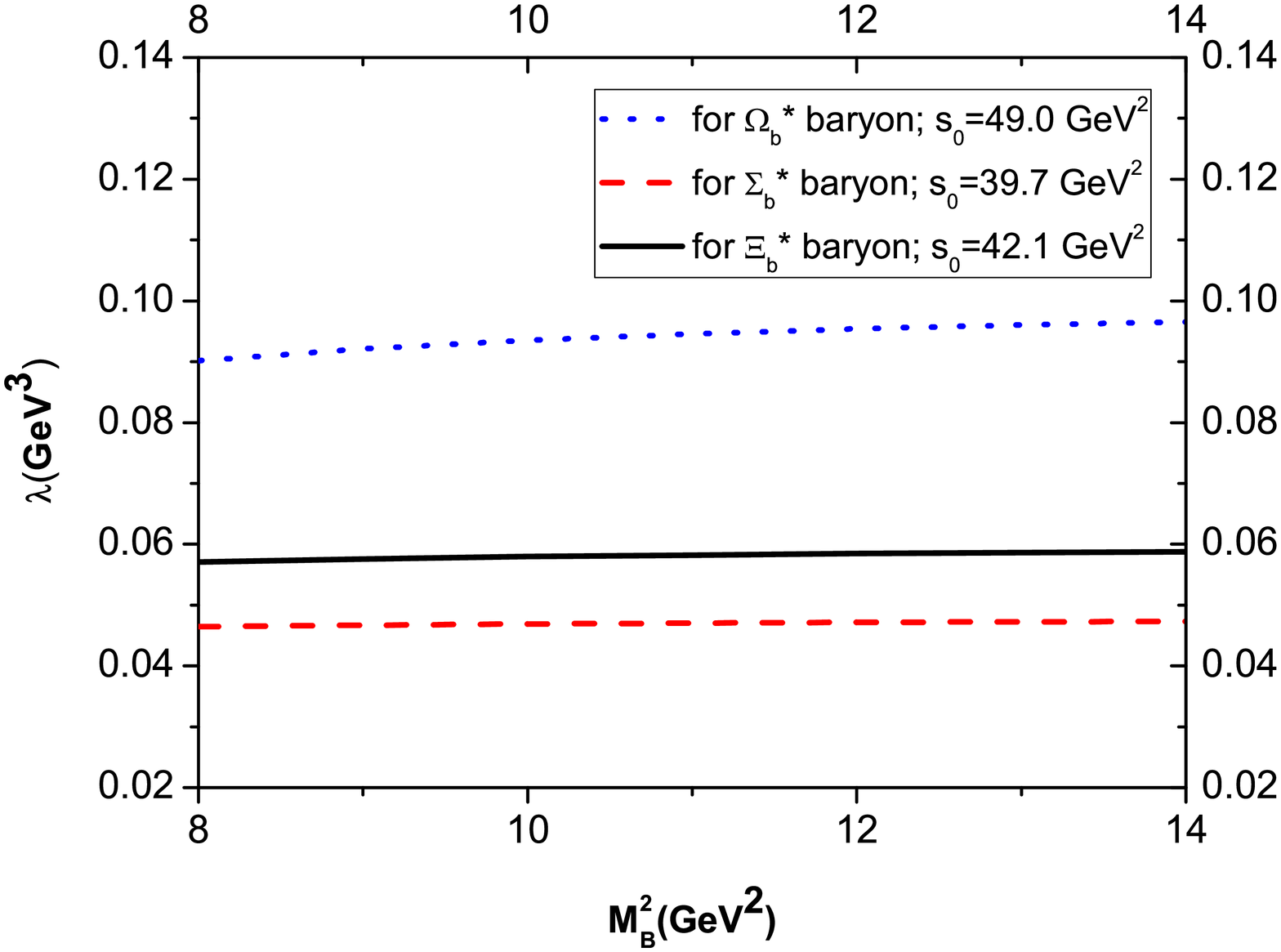}}
\subfigure[]{\includegraphics[width=8cm]{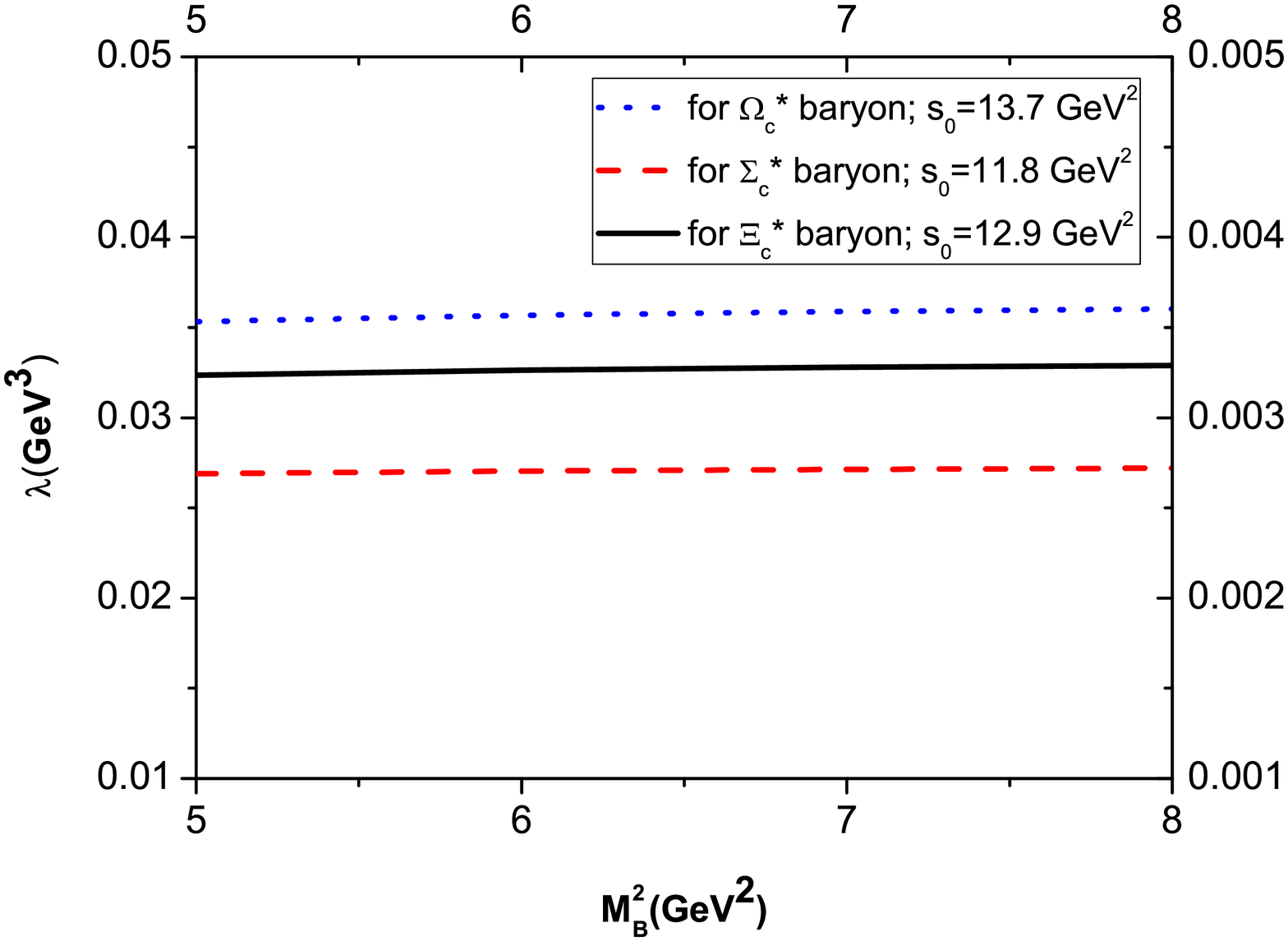}}
\end{center}
\caption{Variations of the mass and residue of the heavy spin--3/2 negative parity baryons  with respect to $M_B^2$ at fixed values of $s_0$.} \label{condd}
\end{figure}
\begin{table}[h]\label{TableMassResidue}
\renewcommand{\arraystretch}{1.5}
\addtolength{\arraycolsep}{3pt}
$$
\begin{array}{|c|c|c|c|c|}
\hline \hline
    \mbox{  Mass and residue   }   & \mbox{Present
    Work}
     &  \cite{Zhi-Gang Wang1}  & \cite{Roberts}  & \cite{Ebert1,Ebert2}
     \\
\hline
 \mbox{$m_{\Omega_b^{*}}~(GeV)$} &  6.40\pm0.26 &6.26\pm0.15  &6.304&6.330 \\
  \hline
   \mbox{$m_{\Omega_c^{*}}~(GeV)$}  &  3.08\pm0.12 &2.98\pm0.16  &2.986&2.998 \\
   \hline
 \mbox{ $m_{\Sigma_b^{*}}~(GeV)$}  &   5.91\pm0.22  &6.00\pm0.18  &6.101& 6.076 \\
  \hline
   \mbox{$m_{\Sigma_c^{*}}~(GeV)$}  &  2.84\pm0.11 & 2.74\pm0.20  &2.763&2.761\\
 \hline
   \mbox{$m_{\Xi_b^{*}}~(GeV)$}  & 6.06\pm0.24  &6.14\pm0.16  &6.194&6.212\\
 \hline
   \mbox{$m_{\Xi_c^{*}}~(GeV)$}  &  2.93\pm0.11 &2.86\pm0.17  &2.871&2.900\\
   \hline
 \mbox{$\lambda_{\Omega_b^{*}}~(GeV^3)$} &0.094\pm0.028 &0.095\pm0.019  &-&- \\
  \hline
   \mbox{$\lambda_{\Omega_c^{*}}~(GeV^3)$}  &0.036\pm0.011 &0.072\pm0.013  &-&- \\
   \hline
 \mbox{ $\lambda_{\Sigma_b^{*}}~(GeV^3)$}  &   0.048\pm0.014  &0.047\pm0.012  &-& - \\
  \hline
   \mbox{$\lambda_{\Sigma_c^{*}}~(GeV^3)$}  &  0.027\pm0.009 & 0.037\pm0.009 &-&-\\
 \hline
   \mbox{$\lambda_{\Xi_b^{*}}~(GeV^3)$}  & 0.058\pm0.017  &0.054\pm0.013  &-&-\\
 \hline
   \mbox{$\lambda_{\Xi_c^{*}}~(GeV^3)$}  &   0.033\pm0.010 &0.045\pm0.009  &-&-\\
                         \hline \hline
\end{array}
$$
\caption{Values of the masses and residues of  the heavy
spin--3/2 negative parity baryons with single heavy bottom or
charm quark.} \label{couplingconstant3}
\renewcommand{\arraystretch}{1}
\addtolength{\arraycolsep}{-1.0pt}
\end{table}
The upper and lower bands on the Borel parameter is determined requiring that not only the contributions of the higher states and continuum are small compared to the pole contributions, but also
the perturbative part exceeds the non-perturbative contributions and the series of sum rules converge (for a discussion on the different but equivalent ways of fixing the Borel parameter see \cite{Straub:2015ica}). By these considerations,
the following working intervals are found:
 \begin{eqnarray}\label{RegionM2}
&&8~GeV^2 \leq M_B^2 \leq 14~GeV^2~\mbox{\rm for $\Omega_b$, $\Sigma_b$
and
$\Xi_b$},\nonumber \\
&&~~~~~~~~~~~~~~~~~~~~~~~~~~\mbox{and}\nonumber \\
&&5~GeV^2 \leq M_B^2 \leq 8~GeV^2~\mbox{\rm for $\Omega_c$,
$\Sigma_c$ and $\Xi_c$}.
\end{eqnarray}
The variations of the  mass and residue of the baryons under consideration with respect to $M_B^2$ at fixed values of $s_0$ are shown in figure 1. From this figure, we see that the mass and residue of these baryons show good stabilities
with respect to the Borel mass parameter in its working regions. Our numerical calculations show also that the dependence of the results on $s_0$ is relatively weak at the above mentioned working region for the continuum
threshold.

We depict the numerical results of the masses and residues of the negative parity heavy spin--3/2 baryons in table \ref{TableMassResidue}. The errors in the presented results are due to the uncertainties in determination of the working regions for the
auxiliary parameters as well as the errors of other input parameters. With a quick glance on this table we read that the values of the masses of the heavy spin--3/2 negative parity baryons obtained in the present work are overall close to the predictions
of \cite{Zhi-Gang Wang1,Roberts,Ebert1,Ebert2}.  For the residue of b-baryons, our results are very close to those of
\cite{Zhi-Gang Wang1}, but for the residues of c-baryons our results considerably small compared to the predictions of \cite{Zhi-Gang Wang1}. The results of the present work for the masses and residues will be used in determination of the
magnetic dipole moments of the negative parity heavy spin--3/2 baryons in the following section.

\section{Magnetic dipole moment of the negative parity spin--3/2 heavy baryons }
In order to obtain the LCSR for  the magnetic dipole moment of the
negative parity heavy spin-3/2 baryons we choose the following two-point correlation function in the presence of a background photon field:
\begin{eqnarray}\label{CorrelationFunction}
\Pi_{\mu\nu}=i \int d^4x~e^{ip\cdot x}~ {\langle}\gamma(q)| {\cal T}\left
( \eta_{\mu}^{{\cal B}}(x)~\bar{\eta}_{\nu}^{{\cal B}}(0)\right)|0{\rangle},
\end{eqnarray}
where $\gamma(q)$ means the electromagnetic field. Note that we use the background electromagnetic plane wave field and the photon DAs instead of the electromagnetic current in accordance with the light cone QCD sum rule approach and the point that at $ q^2=0 $ the em-current gives the charge only. In terms of twist expansion the local em-current is twist-2 but the photon DAs contain also the higher twists.  We will also calculate this correlation function once in terms of hadronic parameters called the physical or hadronic side, and the second in terms of QCD parameters called the QCD or OPE side.
By matching these two sides through a dispersion relation in the Borel scheme one can calculate the magnetic dipole moment of the baryons under consideration in terms of other parameters entering the calculations.

\subsection{Hadronic Side}

For the hadronic side of the calculation one inserts complete sets
of states between the interpolating currents in
Eq.~(\ref{CorrelationFunction}) with quantum numbers of heavy
baryons. Integration over four-$x$ give
\begin{eqnarray}\label{CorrelationFunctionPhySide}
\Pi_{\mu\nu}^{HAD}&=&\frac{{\langle}0|\eta_{\mu}^{{\cal B}}|{\cal
B}^+(p_2){\rangle}{\langle}{\cal B}^+(p_2)\gamma(q)|{\cal
B}^+(p_1){\rangle}{\langle}{\cal
B}^+(p_1)|\bar{\eta}_{\nu}^{{\cal B}}|0{\rangle}}{(m_{{\cal
B}^+}^2-p_1^2)(m_{{\cal B}^+}^2-p_2^2)}
\nonumber \\
&+&\frac{{\langle}0|\eta_{\mu}^{{\cal B}}|{\cal
B}^-(p_2){\rangle}{\langle}{\cal B}^-(p_2)\gamma(q)|{\cal
B}^+(p_1){\rangle}{\langle}{\cal
B}^+(p_1)|\bar{\eta}_{\nu}^{{\cal B}}|0{\rangle}}{(m_{{\cal
B}^+}^2-p_1^2)(m_{{\cal B}^-}^2-p_2^2)}
\nonumber \\
&+&\frac{{\langle}0|\eta_{\mu}^{{\cal B}}|{\cal
B}^+(p_2){\rangle}{\langle}{\cal B}^+(p_2)\gamma(q)|{\cal
B}^-(p_1){\rangle}{\langle}{\cal
B}^-(p_1)|\bar{\eta}_{\nu}^{{\cal B}}|0{\rangle}}{(m_{{\cal
B}^-}^2-p_1^2)(m_{{\cal B}^+}^2-p_2^2)}
\nonumber \\
&+&\frac{{\langle}0|\eta_{\mu}^{{\cal B}}|{\cal
B}^-(p_2){\rangle}{\langle}{\cal B}^-(p_2)\gamma(q)|{\cal
B}^-(p_1){\rangle}{\langle}{\cal
B}^-(p_1)|\bar{\eta}_{\nu}^{{\cal B}}|0{\rangle}}{(m_{{\cal
B}^-}^2-p_1^2)(m_{{\cal B}^-}^2-p_2^2)}
\nonumber \\
&+&\cdots~,
\end{eqnarray}
where $p_1=p+q$, $p_2=p$ and $\cdots$ stands for the contributions coming from the higher
states and continuum. To proceed, beside the matrix elements defined in the previous section in terms of residues, we need the following  matrix elements parametrized in terms of electromagnetic form factors (see also \cite{Aliev1}):
\begin{eqnarray}\label{OtherMatrixElements}
{\langle}{\cal B}^+(p_2)\gamma(q)|{\cal
B}^+(p_1){\rangle}&=&\varepsilon_{\rho}\bar{u}_{\alpha}(p_2)\Bigg\{-g^{\alpha\beta}
\Big[\gamma^{\rho}(f_1
+f_2)+\frac{(p_1+p_2)^{\rho}}{2m_{{\cal{B}}^+}}f_2
+q^{\rho}f_3\Big]
\nonumber \\
&-&\frac{q^{\alpha}q^{\beta}}{2m_{{\cal
B}^+}^2}\Big[\gamma^{\rho}(g_1
+g_2)+\frac{(p_1+p_2)^{\rho}}{2m_{{\cal{B}}^+}}g_2
+q^{\rho}g_3\Big] \Bigg\}u_{\beta}(p_1),
\nonumber \\
{\langle}{\cal B}^-(p_2)\gamma(q)|{\cal
B}^-(p_1){\rangle}&=&\varepsilon_{\rho}\bar{u}_{\alpha}(p_2)\Bigg\{-g^{\alpha\beta}
\Big[\gamma^{\rho}(f_1^*
+f_2^*)+\frac{(p_1+p_2)^{\rho}}{2m_{{\cal{B}}^-}}f_2^*
+q^{\rho}f_3^*\Big]
\nonumber \\
&-&\frac{q^{\alpha}q^{\beta}}{2m_{{\cal
B}^-}^2}\Big[\gamma^{\rho}(g_1^*
+g_2^*)+\frac{(p_1+p_2)^{\rho}}{2m_{{\cal{B}}^-}}g_2^*
+q^{\rho}g_3^* \Big] \Bigg\}u_{\beta}(p_1),
\nonumber \\
 {\langle}{\cal
B}^+(p_2)\gamma(q)|{\cal
B}^-(p_1){\rangle}&=&\varepsilon_{\rho}\bar{u}_{\alpha}(p_2)\Bigg\{-g^{\alpha\beta}
\Big[\gamma^{\rho}(f_1^T
+f_2^T)+\frac{(p_1+p_2)^{\rho}}{m_{{\cal{B}}^+}+m_{{\cal{B}}^-}}f_2^T
+q^{\rho}f_3^T\Big]
\nonumber \\
&-&\frac{q^{\alpha}q^{\beta}}{m_{{\cal B}^+}^2+m_{{\cal
B}^-}^2}\Big[\gamma^{\rho}(g_1^T
+g_2^T)+\frac{(p_1+p_2)^{\rho}}{m_{{\cal{B}}^+}+m_{{\cal{B}}^-}}g_2^T
+q^{\rho}g_3^T\Big] \Bigg\}\gamma_5 u_{\beta}(p_1),
\nonumber \\
 {\langle}{\cal
B}^-(p_2)\gamma(q)|{\cal
B}^+(p_1){\rangle}&=&-\varepsilon_{\rho}\bar{u}_{\alpha}(p_2)\gamma_5\Bigg\{
-g^{\alpha\beta} \Big[\gamma^{\rho}(f_1^{T^*}
+f_2^{T^*})+\frac{(p_1+p_2)^{\rho}}{m_{{\cal{B}}^+}+m_{{\cal{B}}^-}}f_2^{T^*}
+q^{\rho}f_3^{T^*}\Big]
\nonumber \\
&-&\frac{q^{\alpha}q^{\beta}}{m_{{\cal B}^+}^2+m_{{\cal
B}^-}^2}\Big[\gamma^{\rho}(g_1^{T^*}
+g_2^{T^*})+\frac{(p_1+p_2)^{\rho}}{m_{{\cal{B}}^+}+m_{{\cal{B}}^-}}g_2^{T^*}
+q^{\rho}g_3^{T^*}\Big] \Bigg\} u_{\beta}(p_1),
\nonumber \\
\end{eqnarray}
where $f_i$, $f_i^*$,  $f_i^T$,  $f_i^{T^*}$,  $g_i$,  $g_i^*$,
$g_i^T$ and  $g_i^{T^*}$ are electromagnetic  form factors whose values at  $q^2=0$ are needed in determination of the multipole moments. In the
above equation,  $\varepsilon_{\rho}$ is the four-polarization vector of the electromagnetic field.

Here also to remove the pollution from the spin--1/2 baryons  a process similar to that of previous section is followed and
 the ordering $\gamma_\mu \rlap/{\varepsilon} \rlap/{q} \rlap/{p} \gamma_\nu$ is applied.  To complete the task,  the terms containing the  $\gamma_\mu $ at the beginning,
 $\gamma_\nu $ at the end and those proportional to $p_{1\mu}$ and  $p_{2\nu}$  are set to zero.

Substituting  Eqs.~(\ref{MatrixElements})
 and (\ref{OtherMatrixElements}) in
Eq.~(\ref{CorrelationFunctionPhySide}) and using Eq.~(\ref{SpinSummation}) the final form of the hadronic side of the correlation
function in Borel scheme is obtained as
\begin{eqnarray}\label{CorrelationFunctionLast}
\widehat{\textbf{B}}\Pi_{\mu\nu}^{HAD}&=&\Bigg[\lambda_{{\cal{B}^-}}
\lambda_{{\cal{B}^+}}m_{{\cal{B}^-}}\Big
(f_1^T+f_2^T\Big)e^{-\frac{m_{{\cal
{B}^-}}^2}{M_1^2}-\frac{m_{{\cal {B}^+}}^2}{M_2^2}}
+\lambda_{{\cal{B}^-}}\lambda_{{\cal{B}^+}}m_{{\cal{B}^+}}\Big
(f_1^{T^*}+f_2^{T^*}\Big)e^{-\frac{m_{{\cal
{B}^+}}^2}{M_1^2}-\frac{m_{{\cal {B}^-}}^2}{M_2^2}}
\nonumber \\
&-&\lambda_{{\cal{B}^-}}^2m_{{\cal{B}^-}}\Big(f_1^{*}+f_2^{*}\Big)e^{-\frac{m_{{\cal
{B}^-}}^2}{M_1^2}-\frac{m_{{\cal
{B}^-}}^2}{M_2^2}}-\lambda_{{\cal{B}^+}}^2m_{{\cal{B}^+}}(f_1+f_2\Big)
e^{-\frac{m_{{\cal {B}^+}}^2}{M_1^2}-\frac{m_{{\cal
{B}^+}}^2}{M_2^2}}
 \Bigg] g_{\mu\nu}\not\!\varepsilon \not\!q
\nonumber \\
&+&\Bigg[\lambda_{{\cal{B}^-}}^2\Big(m_{{\cal{B}^-}}^2-m_{{\cal{B}^+}}^2\Big)
\Big(f_1^{*}+f_2^{*}\Big)e^{-\frac{m_{{\cal
{B}^-}}^2}{M_1^2}-\frac{m_{{\cal
{B}^-}}^2}{M_2^2}}+\lambda_{{\cal{B}^-}}\lambda_{{\cal{B}^+}}m_{{\cal{B}^-}}
\Big(m_{{\cal{B}^-}}+m_{{\cal{B}^+}}\Big)\Big(f_1^{T}+f_2^{T}\Big)
\nonumber \\
&\times& e^{-\frac{m_{{\cal {B}^-}}^2}{M_1^2}-\frac{m_{{\cal
{B}^+}}^2}{M_2^2}}-\lambda_{{\cal{B}^-}}\lambda_{{\cal{B}^+}}m_{{\cal{B}^+}}
\Big(m_{{\cal{B}^+}}+m_{{\cal{B}^-}}\Big)\Big(f_1^{T^*}+f_2^{T^*}\Big)
e^{-\frac{m_{{\cal {B}^+}}^2}{M_1^2}-\frac{m_{{\cal
{B}^-}}^2}{M_2^2}} \Bigg] g_{\mu\nu}\not\!\varepsilon
\nonumber \\
&-&\Bigg[\lambda_{{\cal{B}^-}}\lambda_{{\cal{B}^+}}\Big(f_1^{T}+f_2^{T}\Big)e^{-\frac{m_{{\cal
{B}^-}}^2}{M_1^2}-\frac{m_{{\cal
{B}^+}}^2}{M_2^2}}-\lambda_{{\cal{B}^-}}\lambda_{{\cal{B}^+}}
\Big(f_1^{T^*}+f_2^{T^*}\Big)e^{-\frac{m_{{\cal
{B}^+}}^2}{M_1^2}-\frac{m_{{\cal {B}^-}}^2}{M_2^2}}
\nonumber \\
&+&\lambda_{{\cal{B}^+}}^2\Big(f_1+f_2\Big)e^{-\frac{m_{{\cal
{B}^+}}^2}{M_1^2}-\frac{m_{{\cal
{B}^+}}^2}{M_2^2}}-\lambda_{{\cal{B}^-}}^2\Big(f_1^{*}+f_2^{*}\Big)e^{-\frac{m_{{\cal
{B}^-}}^2}{M_1^2}-\frac{m_{{\cal {B}^-}}^2}{M_2^2}}
 \Bigg] g_{\mu\nu}\not\!p \not\!\varepsilon \not\!q
\nonumber \\
&-&\Bigg[
\lambda_{{\cal{B}^-}}\lambda_{{\cal{B}^+}}\Big(m_{{\cal{B}^-}}+m_{{\cal{B}^+}}\Big)
\Big(f_1^{T^*}+f_2^{T^*}\Big) e^{-\frac{m_{{\cal
{B}^+}}^2}{M_1^2}-\frac{m_{{\cal
{B}^-}}^2}{M_2^2}}+\lambda_{{\cal{B}^-}}\lambda_{{\cal{B}^+}}
\Big(m_{{\cal{B}^-}}+m_{{\cal{B}^+}}\Big)
\nonumber \\
&\times& \Big(f_1^{T}+f_2^{T}\Big) e^{-\frac{m_{{\cal
{B}^-}}^2}{M_1^2}-\frac{m_{{\cal {B}^+}}^2}{M_2^2}}
 \Bigg] g_{\mu\nu}\not\!p \not\!\varepsilon+ \mbox{other
 structures}.
\end{eqnarray}
where $M_1^2$, $M_2^2$ are the Borel mass parameters in the initial and final channels, respectively. We are interested in calculation of the magnetic dipole moment of only the negative parity baryons
 which is given as
$\mu=3(f_1^*+f_2^*)$ in the unit of their natural magneton, i.e. $e\hbar/(2m_{{\cal{B}}}c)$. So we have
diagonal transitions for this case and the initial and final baryon masses are the same. Hence, in
the calculations, we take $M_1^2=M_2^2=M_B^2=2M^2$.

\subsection{OPE Side}

In order to obtain the OPE side of the correlation function, we
insert the interpolating current of the heavy spin--3/2 baryons   into
Eq.~(\ref{CorrelationFunction}). After performing  contractions of all
quark pairs using the Wick's theorem, we get
\begin{eqnarray}\label{QCDSide}
\Pi_{\mu\nu}^{OPE}&=&-iA^{2}\epsilon_{abc}\epsilon_{a'b'c'}\int
d^{4}xe^{ipx}\langle\gamma(q)\mid\{S_{Q}^{ca'}
\gamma_{\nu}S'^{bb'}_{q_{2}}\gamma_{\mu}S_{q_{1}}^{ac'}\nonumber\\&+&S_{Q}^{cb'}
\gamma_{\nu}S'^{aa'}_{q_{1}}\gamma_{\mu}S_{q_{2}}^{bc'}+
S_{q_{2}}^{ca'}
\gamma_{\nu}S'^{bb'}_{q_{1}}\gamma_{\mu}S_{Q}^{ac'}+S_{q_{2}}^{cb'}
\gamma_{\nu}S'^{aa'}_{Q}\gamma_{\mu}S_{q_{1}}^{bc'}\nonumber\\&+&
S_{q_{1}}^{cb'}
\gamma_{\nu}S'^{aa'}_{q_{2}}\gamma_{\mu}S_{Q}^{bc'}+S_{q_{1}}^{ca'}
\gamma_{\nu}S'^{bb'}_{Q}\gamma_{\mu}S_{q_{2}}^{ac'}
+Tr(\gamma_{\mu}S_{q_{1}}^{ab'}\gamma_{\nu}S'^{ba'}_{q_{2}})S^{cc'}_{Q}\nonumber\\&+&Tr(\gamma_{\mu}S_{Q}^{ab'}\gamma_{\nu}S'^{ba'}_{q_{1}})S^{cc'}_{q_{2}}+Tr(\gamma_{\mu}S_{q_{2}}^{ab'}\gamma_{\nu}S'^{ba'}_{Q})S^{cc'}_{q_{1}}\}\mid
0\rangle,
\end{eqnarray}
where $S'=CS^TC$; and $S_Q$ and $S_q$ are the heavy and light quark
propagators, respectively. The correlation function in OPE side
contains three different contributions:
\begin{itemize}
    \item perturbative contributions,
    \item mixed contributions where the photon is radiated from the short distances and at least one of
the quarks interacts with the QCD vacuum and makes a condensate and
    \item non-perturbative contributions where the photon is radiated
at long distances.
\end{itemize}
To proceed in calculations of these contributions, we use the heavy and light quark propagators again in coordinate space.
 We also need some matrix elements defined in terms of   the photon DAs  \cite {Ball}:
 \begin{eqnarray}
&&\langle \gamma(q) \vert  \bar q(x) \sigma_{\mu \nu} q(0) \vert 0
\rangle  = -i e_q \bar q q (\varepsilon_\mu q_\nu -
\varepsilon_\nu q_\mu) \int_0^1 du e^{i \bar u qx} \left(\chi
\varphi_\gamma(u) + \frac{x^2}{16} \mathbb{A}  (u) \right)
\nonumber \\ && -\frac{i}{2(qx)}  e_q \qq \left[x_\nu
\left(\varepsilon_\mu - q_\mu \frac{\varepsilon x}{qx}\right) -
x_\mu \left(\varepsilon_\nu - q_\nu \frac{\varepsilon x}{q
x}\right) \right] \int_0^1 du e^{i \bar u q x} h_\gamma(u),
\nonumber \\
&&\langle \gamma(q) \vert  \bar q(x) \gamma_\mu q(0) \vert 0
\rangle = e_q f_{3 \gamma} \left(\varepsilon_\mu - q_\mu
\frac{\varepsilon x}{q x} \right) \int_0^1 du e^{i \bar u q x}
\psi^v(u),
\nonumber \\
&&\langle \gamma(q) \vert \bar q(x) \gamma_\mu \gamma_5 q(0) \vert
0 \rangle  = - \frac{1}{4} e_q f_{3 \gamma} \epsilon_{\mu \nu
\alpha \beta } \varepsilon^\nu q^\alpha x^\beta \int_0^1 du e^{i
\bar u q x} \psi^a(u),
\nonumber \\
&&\langle \gamma(q) | \bar q(x) g_s G_{\mu \nu} (v x) q(0) \vert 0
\rangle = -i e_q \qq \left(\varepsilon_\mu q_\nu - \varepsilon_\nu
q_\mu \right) \int {\cal D}\alpha_i e^{i (\alpha_{\bar q} + v
\alpha_g) q x} {\cal S}(\alpha_i),
\nonumber \\
&&\langle \gamma(q) | \bar q(x) g_s \tilde G_{\mu \nu} i \gamma_5
(v x) q(0) \vert 0 \rangle = -i e_q \qq \left(\varepsilon_\mu
q_\nu - \varepsilon_\nu q_\mu \right) \int {\cal D}\alpha_i e^{i
(\alpha_{\bar q} + v \alpha_g) q x} \tilde {\cal S}(\alpha_i),
\nonumber \\
&&\langle \gamma(q) \vert \bar q(x) g_s \tilde G_{\mu \nu}(v x)
\gamma_\alpha \gamma_5 q(0) \vert 0 \rangle = e_q f_{3 \gamma}
q_\alpha (\varepsilon_\mu q_\nu - \varepsilon_\nu q_\mu) \int
{\cal D}\alpha_i e^{i (\alpha_{\bar q} + v \alpha_g) q x} {\cal
A}(\alpha_i),
\nonumber \\
&&\langle \gamma(q) \vert \bar q(x) g_s G_{\mu \nu}(v x) i
\gamma_\alpha q(0) \vert 0 \rangle = e_q f_{3 \gamma} q_\alpha
(\varepsilon_\mu q_\nu - \varepsilon_\nu q_\mu) \int {\cal
D}\alpha_i e^{i (\alpha_{\bar q} + v \alpha_g) q x} {\cal
V}(\alpha_i) ,\nonumber \\ && \langle \gamma(q) \vert \bar q(x)
\sigma_{\alpha \beta} g_s G_{\mu \nu}(v x) q(0) \vert 0 \rangle  =
e_q \qq \left\{
        \left[\left(\varepsilon_\mu - q_\mu \frac{\varepsilon x}{q x}\right)\left(g_{\alpha \nu} -
        \frac{1}{qx} (q_\alpha x_\nu + q_\nu x_\alpha)\right) \right. \right. q_\beta
\nonumber \\ && -
         \left(\varepsilon_\mu - q_\mu \frac{\varepsilon x}{q x}\right)\left(g_{\beta \nu} -
        \frac{1}{qx} (q_\beta x_\nu + q_\nu x_\beta)\right) q_\alpha
\nonumber \\ && -
         \left(\varepsilon_\nu - q_\nu \frac{\varepsilon x}{q x}\right)\left(g_{\alpha \mu} -
        \frac{1}{qx} (q_\alpha x_\mu + q_\mu x_\alpha)\right) q_\beta
\nonumber \\ &&+
         \left. \left(\varepsilon_\nu - q_\nu \frac{\varepsilon x}{q.x}\right)\left( g_{\beta \mu} -
        \frac{1}{qx} (q_\beta x_\mu + q_\mu x_\beta)\right) q_\alpha \right]
   \int {\cal D}\alpha_i e^{i (\alpha_{\bar q} + v \alpha_g) qx} {\cal T}_1(\alpha_i)
\nonumber \\ &&+
        \left[\left(\varepsilon_\alpha - q_\alpha \frac{\varepsilon x}{qx}\right)
        \left(g_{\mu \beta} - \frac{1}{qx}(q_\mu x_\beta + q_\beta x_\mu)\right) \right. q_\nu
\nonumber \\ &&-
         \left(\varepsilon_\alpha - q_\alpha \frac{\varepsilon x}{qx}\right)
        \left(g_{\nu \beta} - \frac{1}{qx}(q_\nu x_\beta + q_\beta x_\nu)\right)  q_\mu
\nonumber \\ && -
         \left(\varepsilon_\beta - q_\beta \frac{\varepsilon x}{qx}\right)
        \left(g_{\mu \alpha} - \frac{1}{qx}(q_\mu x_\alpha + q_\alpha x_\mu)\right) q_\nu
\nonumber \\ &&+
         \left. \left(\varepsilon_\beta - q_\beta \frac{\varepsilon x}{qx}\right)
        \left(g_{\nu \alpha} - \frac{1}{qx}(q_\nu x_\alpha + q_\alpha x_\nu) \right) q_\mu
        \right]
    \int {\cal D} \alpha_i e^{i (\alpha_{\bar q} + v \alpha_g) qx} {\cal T}_2(\alpha_i)
\nonumber \\ &&+
        \frac{1}{qx} (q_\mu x_\nu - q_\nu x_\mu)
        (\varepsilon_\alpha q_\beta - \varepsilon_\beta q_\alpha)
    \int {\cal D} \alpha_i e^{i (\alpha_{\bar q} + v \alpha_g) qx} {\cal T}_3(\alpha_i)
\nonumber \\ &&+
        \left. \frac{1}{qx} (q_\alpha x_\beta - q_\beta x_\alpha)
        (\varepsilon_\mu q_\nu - \varepsilon_\nu q_\mu)
    \int {\cal D} \alpha_i e^{i (\alpha_{\bar q} + v \alpha_g) qx} {\cal T}_4(\alpha_i)
                        \right\},
\end{eqnarray}
where $\varphi_\gamma(u)$ is the leading twist-2 photon DAs,  $\psi^v(u)$,
$\psi^a(u)$, ${\cal A}(\alpha_i)$ and ${\cal V}(\alpha_i)$ are twist 3 and
$h_\gamma(u)$, $\mathbb{A}$(u) and ${\cal T}_i$ ($i=1,~2,~3,~4$) are twist 4 photon DAs \cite{Ball}.  In the above equations  $\chi$ is the magnetic
susceptibility of the light quarks. The measure $\int{\cal D} \alpha_i$  is defined as
\begin{equation}
\int {\cal D} \alpha_i = \int_0^1 d \alpha_{\bar q} \int_0^1 d
\alpha_q \int_0^1 d \alpha_g \delta(1-\alpha_{\bar
q}-\alpha_q-\alpha_g),\nonumber \\
\end{equation}
and  the  photon DAs are given as \cite{Ball}:
\begin{eqnarray}
\varphi_\gamma(u) &=& 6 u \bar u \left( 1 + \varphi_2(\mu)
C_2^{\frac{3}{2}}(u - \bar u) \right),
\nonumber \\
\psi^v(u) &=& 3 \left(3 (2 u - 1)^2 -1 \right)+\frac{3}{64}
\left(15 w^V_\gamma - 5 w^A_\gamma\right)
                        \left(3 - 30 (2 u - 1)^2 + 35 (2 u -1)^4
                        \right),
\nonumber \\
\psi^a(u) &=& \left(1- (2 u -1)^2\right)\left(5 (2 u -1)^2
-1\right) \frac{5}{2}
    \left(1 + \frac{9}{16} w^V_\gamma - \frac{3}{16} w^A_\gamma
    \right),
\nonumber \\
{\cal A}(\alpha_i) &=& 360 \alpha_q \alpha_{\bar q} \alpha_g^2
        \left(1 + w^A_\gamma \frac{1}{2} (7 \alpha_g - 3)\right),
\nonumber \\
{\cal V}(\alpha_i) &=& 540 w^V_\gamma (\alpha_q - \alpha_{\bar q})
\alpha_q \alpha_{\bar q}
                \alpha_g^2,
\nonumber \\
h_\gamma(u) &=& - 10 \left(1 + 2 \kappa^+\right)
C_2^{\frac{1}{2}}(u - \bar u),
\nonumber \\
\mathbb{A}(u) &=& 40 u^2 \bar u^2 \left(3 \kappa - \kappa^+
+1\right) \nonumber \\ && +
        8 (\zeta_2^+ - 3 \zeta_2) \left[u \bar u (2 + 13 u \bar u) \right.
\nonumber \\ && + \left.
                2 u^3 (10 -15 u + 6 u^2) \ln(u) + 2 \bar u^3 (10 - 15 \bar u + 6 \bar u^2)
        \ln(\bar u) \right],
\nonumber \\
{\cal T}_1(\alpha_i) &=& -120 (3 \zeta_2 + \zeta_2^+)(\alpha_{\bar
q} - \alpha_q)
        \alpha_{\bar q} \alpha_q \alpha_g,
\nonumber \\
{\cal T}_2(\alpha_i) &=& 30 \alpha_g^2 (\alpha_{\bar q} -
\alpha_q)
    \left((\kappa - \kappa^+) + (\zeta_1 - \zeta_1^+)(1 - 2\alpha_g) +
    \zeta_2 (3 - 4 \alpha_g)\right),
\nonumber \\
{\cal T}_3(\alpha_i) &=& - 120 (3 \zeta_2 -
\zeta_2^+)(\alpha_{\bar q} -\alpha_q)
        \alpha_{\bar q} \alpha_q \alpha_g,
\nonumber \\
{\cal T}_4(\alpha_i) &=& 30 \alpha_g^2 (\alpha_{\bar q} -
\alpha_q)
    \left((\kappa + \kappa^+) + (\zeta_1 + \zeta_1^+)(1 - 2\alpha_g) +
    \zeta_2 (3 - 4 \alpha_g)\right),\nonumber \\
{\cal S}(\alpha_i) &=& 30\alpha_g^2\{(\kappa +
\kappa^+)(1-\alpha_g)+(\zeta_1 + \zeta_1^+)(1 - \alpha_g)(1 -
2\alpha_g)\nonumber \\&+&\zeta_2
[3 (\alpha_{\bar q} - \alpha_q)^2-\alpha_g(1 - \alpha_g)]\},\nonumber \\
\tilde {\cal S}(\alpha_i) &=&-30\alpha_g^2\{(\kappa -
\kappa^+)(1-\alpha_g)+(\zeta_1 - \zeta_1^+)(1 - \alpha_g)(1 -
2\alpha_g)\nonumber \\&+&\zeta_2 [3 (\alpha_{\bar q} -
\alpha_q)^2-\alpha_g(1 - \alpha_g)]\},
\end{eqnarray}
where
 $\varphi_2(1~GeV) = 0$, $w^V_\gamma = 3.8 \pm 1.8$,
$w^A_\gamma = -2.1 \pm 1.0$, $\kappa = 0.2$, $\kappa^+ = 0$,
$\zeta_1 = 0.4$, $\zeta_2 = 0.3$, $\zeta_1^+ = 0$ and $\zeta_2^+ =
0$  \cite{Ball}.

After lengthy calculations (for details see for instance \cite{hayriye,altug}), the OPE side of the correlation function is
obtained in terms of the selected structures as
\begin{eqnarray}\label{QCDSide}
\widehat{\textbf{B}}\Pi_{\mu\nu}^{OPE}&=&\Pi_1^{OPE}
g_{\mu\nu}\not\!\varepsilon
\not\!q+\Pi_2^{OPE}g_{\mu\nu}\not\!\varepsilon+\Pi_3^{OPE}
g_{\mu\nu}\not\!p \not\!\varepsilon \not\!q +
\Pi_4^{OPE}g_{\mu\nu}\not\!p \not\!\varepsilon+ \mbox{other
 structures},\nonumber \\
 \end{eqnarray}
where  $ \Pi_{1,2,3,4}^{OPE} $ are very lengthy functions, hence we do not present their explicit expressions here.

Having obtained both the hadronic and OPE sides of the correlation
function in Borel scheme it is the time for equating the two sides
in order to obtain LCSR for the magnetic moment of the
spin-$3/2^-$ baryons with single heavy bottom or charm quark. Before doing this we should remind that the magnetic dipole moment is defined in terms of form 
factors, in the unit of the natural magneton, i.e. $e\hbar/(2m_{{\cal{B}}}c)$, as $\mu=3(f_1^*+f_2^*)$ at $q^2=0$, where the factor $3$ is due the fact that
in the non-relativistic limit the interaction Hamiltonian with magnetic field
is given as $\mu B=3(f_1^*+f_2^*)B$. After replacement $f_1^*+f_2^*\rightarrow \mu/3$ in the final expression of the physical side in Eq. (\ref{CorrelationFunctionLast}) and equating the obtained result
 to the OPE side in Eq. (\ref{QCDSide}), we get
the following expression for the magnetic dipole moment of the baryons under consideration:
\begin{eqnarray}\label{MagMoment}
\mu=3\frac{\Pi^{OPE}_2-(m_{{\cal{B}}^+}+m_{{\cal{B}}^-})\Pi^{OPE}_1+m_{{\cal{B}}^+}
(m_{{\cal{B}}^+}+m_{{\cal{B}}^-})\Pi^{OPE}_3-m_{{\cal{B}}^+}\Pi^{OPE}_4}{2\lambda_{{\cal{B}^-}}^2
m_{{\cal{B}}^-}
(m_{{\cal{B}}^+}+m_{{\cal{B}}^-})}e^{\frac{m_{{\cal{B}}^-}^2}{M^2}}.
\end{eqnarray}

The numerical values for the  magnetic dipole moments of the heavy  spin--3/2 negative parity baryons in units of nuclear magneton are presented in
table 3.
\begin{table}[h]\label{TableMagMom}
\renewcommand{\arraystretch}{1.5}
\addtolength{\arraycolsep}{3pt}
$$
\begin{array}{|c|c|}
\hline \hline
    \mbox{  Heavy baryons with $J^{P}=\frac{3}{2}^-$ }  & \mbox{$\mu$}
     \\
\hline
 \mbox{$\Omega_b^{*^-}$} & -1.37\pm0.41 \\
  \hline
   \mbox{$\Omega_c^{*^0}$}  &  -3.46\pm1.04  \\
   \hline
 \mbox{ $\Sigma_b^{*^0}$}  &  0.55\pm0.17   \\
  \hline
   \mbox{$\Sigma_c^{*^+}$}  &  2.10\pm0.59 \\
 \hline
   \mbox{$\Sigma_b^{*^+}$}  & 2.24\pm0.67 \\
 \hline
   \mbox{$\Sigma_c^{*^{^{++}}}$}  &   7.73\pm2.31 \\
   \hline
 \mbox{$\Sigma_b^{*^-}$} &  -1.53\pm0.46  \\
  \hline
   \mbox{$\Sigma_c^{*^0}$}  & -4.02\pm1.16 \\
   \hline
 \mbox{ $\Xi_b^{*^0}$}  &   0.57\pm0.17 \\
  \hline
   \mbox{$\Xi_c^{*^+}$}  &2.29\pm0.68 \\
 \hline
   \mbox{$\Xi_b^{*^-}$}  & -1.32\pm0.39 \\
 \hline
   \mbox{$\Xi_c^{*^0}$}  &  -2.95\pm0.83\\
                         \hline \hline
\end{array}
$$

\caption{Values of the magnetic moment of  the heavy spin-$3/2$
negative parity baryons with single heavy bottom or charm quark
in units of nuclear magneton $\mu_N$. The signs $+, - $ and $0 $  on baryons show their charge.} \label{couplingconstant3}
\renewcommand{\arraystretch}{1}
\addtolength{\arraycolsep}{-1.0pt}
\end{table}
The errors in numerical values in the presented results again belong to the uncertainties in calculations of the working regions for the Borel mass parameter $M^2$ and the continuum threshold
$s_0$, those uncertainties coming from the parameters entering the photon DAs as well as the uncertainties of other input parameters. Quantitatively, in average,  $47\%$, $14\%$ and $39\%$ of the
uncertainties belong to the variations of $s_0$, $M^2$ and DAs together with other inputs, respectively.  When we compare these results with the magnetic dipole moments of the
positive parity spin--3/2 heavy baryons \cite{altug}, we see that the magnetic dipole moments of the negative parity heavy baryons are compatible with those of the positive parity baryons except for
the   $\Omega_c^{*^0}$, $\Sigma_c^{*^0}$,  $\Sigma_c^{*^{^{++}}}$ and $\Xi_c^{*^0}$  baryons which there exist considerable differences in the values. Hence the naive expectation, relation  between the magnetic
moments and masses of the positive and negative parity baryons, i.e., 
\begin{eqnarray}
 \mu_{\mbox{n }}=\frac{m_{\mbox{p}}}{m_{\mbox{n}}}\mu_{\mbox{p }},
\end{eqnarray}
where n and p stand for the negative and positive parity heavy baryons respectively, holds for all $b$-baryons and some of $c$-baryons, but is considerably violated for  the  $c$-baryons mentioned above.
This violation can be attributed to the fact that in our calculations we take into account also the contributions of the  positive-to-positive,  positive-to-negative and negative-to-positive
transitions that affect the $c$-baryons more compared to the $b$-baryons.
 The sign of the magnetic dipole moments of the heavy spin--3/2 baryons with both parities are the same.
Our results may be checked via other non-perturbative approaches as well as by
future experiments.

\section{Conclusion}
We calculated the masses and residues of the negative parity heavy spin--3/2 baryons with single heavy  $b$ or $c$ quark in the framework of QCD sum rules and compared the results
with the existing predictions in the literature. We used the values obtained to calculate the electromagnetic form factors and
finally the values of the magnetic dipole moments of the considered baryons in the context of the light cone QCD sum rules using the photon DAs. Our results may be checked via different non-perturbative methods.
Checking our predictions by future experiments can be very useful for understanding the internal structure as well as the  geometric shape  of the negative parity heavy spin--3/2 baryons.



\begin{thebibliography}{99}
\bibitem{bra} V. M. Braun, A. Lenz, M. Wittmann, Phys. Rev. D 73,  094019 (2006).
\bibitem{Jefferson} V. Punjabi et al., Phys. Rev. C 71, 055202 (2005).
\bibitem{Jones} M. K. Jones et al. [Jefferson Lab Hall A Collaboration], Phys. Rev. Lett. 84, 1398 (2000).
\bibitem{Gayou1} O. Gayou et al., Phys. Rev. C 64,  038202 (2001).
\bibitem{Gayou2}  O. Gayou et al. [Jefferson Lab Hall A Collaboration], Phys. Rev. Lett. 88, 092301 (2002). 
\bibitem{Walker} R. C. Walker et al., Phys. Rev. D 49, 5671 (1994).
\bibitem{Andivahis}  L. Andivahis et al., Phys. Rev. D 50, 5491 (1994).
\bibitem{Litt}  J. Litt et al., Phys. Lett. B 31, 40 (1970).
\bibitem{Berger}  C. Berger, V. Burkert, G. Knop, B. Langenbeck and K. Rith, Phys. Lett. B 35, 87  (1971).
\bibitem{Janssens}  T. Janssens, R. Hofstadter, E. B. Huges and M. R. Yearian, Phys. Rev. 142, 922 (1966).
\bibitem{Arrington} J. Arrington, Phys. Rev. C 68, 034325 (2003).
\bibitem{Christy}  M. E. Christy et al. [E94110 Collaboration], Phys. Rev. C 70, 015206 (2004).
\bibitem{Qattan}  I. A. Qattan et al., Phys. Rev. Lett. 94, 142301 (2005).



\bibitem{Mainz3} M. Kotulla, Prog. Part. Nucl. Phys. 61, 147 (2008).

\bibitem{Chung} Y. Chung, H. G. Dosch, M. Kremer and D. Schall, Nucl. Phys. B 197, 55 (1982).
\bibitem{Oka} M. Oka, D. Jido, A. Hosaka, Nucl. Phys. A 629, 156 (1998).
\bibitem{Kondo} Y. Kondo, O. Morimatsu,  T. Nishikawa,  Nucl. Phys. A 764, 303 (2006).



\bibitem{Aliev}  T. M. Aliev, M. Savci, Phys. Rev. D 89, 053003 (2014).

\bibitem{savci3} T. M. Aliev, M. Savci, J. Phys. G 41, 075007  (2014).

\bibitem{hayriye} K. Azizi, H. Sundu,  Phys. Rev. D 91, 093012  (2015).
\bibitem{Zhi-Gang Wang1} Zhi-Gang Wang, Eur. Phys. J. A 47, 81 (2011).
\bibitem{Zhi-Gang Wang2} Zhi-Gang Wang, Eur. Phys. J. C 68, 479 (2010).
\bibitem{Roberts} W. Roberts and M. Pervin, Int. J. Mod. Phys. A23, 2817 (2008).

\bibitem{Ebert1}D. Ebert, R. N. Faustov and V. O. Galkin, Phys. Lett. B659 (2008)
612.

\bibitem{Ebert2}D. Ebert, R. N. Faustov, V. O. Galkin and A. P. Martynenko, Phys.
Rev. D66, 014008 (2002).

\bibitem{savci4} T. M. Aliev, K. Azizi, M. Savci, arXiv:1504.00172 [hep-ph].
\bibitem{savci5}  T. M. Aliev, T. Barakat, M. Savci, arXiv:1504.08187 [hep-ph].
\bibitem{savci6} T. M. Aliev, K. Azizi, T. Barakat, M. Savci, arXiv:1505.07977 [hep-ph].
\bibitem{Shifman} M. A. Shifman, A. I. Vainshtein and V. I. Zakharov, Nucl. Phys. B 147, 385 (1979); Nucl. Phys. B 147, 448 (1979).



\bibitem{altug} T. M. Aliev, K. Azizi, A. Ozpineci, Nucl. Phys. B 808, 137 (2009).

\bibitem{Savvidy} K. G. Savvidy, arXiv:1005.3455 [hep-th].

\bibitem{Aliev1} T. M. Aliev, M. Savci, Phys.Rev. D90, 116006, 11
(2014).

\bibitem{Ioffe}  B. L. Ioffe, Nuclear Physics B 188, 317 (1981).
\bibitem{Dosch}  H. G. Dosch, M. Jamin and S. Narison, Phys. Lett. B 220, 251 (1989).
 \bibitem{Lee} F. X. Lee,  Phys.Rev. D 57, 1801 (1998).

\bibitem{Balitsky} I. I.  Balitsky,  V. M.  Braun,  Nucl. Phys.  B  311, 541 (1989).


\bibitem{belyaev}  V. M. Belyaev, B. L. Ioffe, Sov. Phys. JETP 57, 716 (1983);
 E. Bagan, M. Chabab, H. G. Dosch and S. Narison, Phys. Lett. B 287, 176 (1992).

\bibitem{Straub:2015ica} 
  A.~Bharucha, D.~M.~Straub and R.~Zwicky,
  JHEP  1608, 098 (2016).







\bibitem{Ball} P.  Ball,  V. M.  Braun, N. Kivel,  Nucl. Phys.  B  649, 263 (2003).









\end{thebibliography}
\end{document}